\documentclass[print]{aa} % for a referee version
\usepackage{graphicx}
\usepackage{txfonts}
\usepackage{natbib}
\bibpunct{(}{)}{;}{a}{}{,}
\begin{document}
   \title{A third cluster of red supergiants in the vicinity of the massive cluster RSGC3\thanks{Based on observations collected at the European Organisation for Astronomical Research in the Southern Hemisphere, Chile, under proposal 087.D-0413A}}

   \author{C. Gonz\'{a}lez-Fern\'{a}ndez\inst{1}
          \and
          I. Negueruela\inst{1}
          }

   \institute{Departamento de F\'{\i}sica, Ingenier\'{\i}a de Sistemas y Teor\'{\i}a de la Se\~{n}al, Universidad de Alicante, Apdo. 99, E03080 Alicante, Spain\\
              \email{carlos.gonzalez@ua.es}
             }

   \date{Accepted for A\& A on January 16, 2012}

% \abstract{}{}{}{}{} 
% 5 {} token are mandatory
 
  \abstract
   {
Recent studies have shown that the area around the massive, obscured cluster RSGC3 may harbour several clusters of red supergiants.
	}
   {
	We analyse a clump of photometrically selected red supergiant candidates 20' south of RSGC3 in order to confirm the existence of another of these clusters.
	}
   {
   Using medium-resolution infrared spectroscopy around $2.27~\mathrm{\mu m}$, we derived spectral types and velocities along the line of sight for the selected candidates, confirming their nature and possible association.
   }
   {
   We find a compact clump of eight red supergiants and four other candidates at some distance, all of them spectroscopically confirmed red supergiants. The majority of these objects must form an open cluster, which we name Alicante 10. Because of the high reddening and strong field contamination, the cluster sequence is not clearly seen in 2MASS or GPS-UKIDSS. From the observed sources, we derive $E(J-K_{\mathrm{S}})=2.6$ and $d\approx6~\mathrm{kpc}$.
   }
   {Although the cluster is smaller than RSGC3, it has an initial mass in excess of $10~000~M_{\odot}$, and it seems to be part of the RSGC3 complex. With the new members this association already has 35 spectroscopically confirmed red supergiants, confirming its place as one of the most active sites of recent stellar formation in the Galaxy.}

	\titlerunning{A third cluster in the vicinity of RSGC3}
	\maketitle
%
%________________________________________________________________

\section{Introduction}

   Red supergiants (RSG) are massive stars that have already ended their hydrogen core-burning phase, and now occupy the reddest part of the Hertszprung-Russell (HR) diagram. Having roughly the same bolometric luminosity as their blue progenitors, with $\log(L_{\mathrm{bol}}/L_\odot)\sim4.0-5.8$ \citep{MM00}, they are extremely luminous sources in the infrared, and this makes them detectable even under heavy interstellar obscuration.
   
   It is this property that has facilitated the discovery of several clusters of RSGs in the region of the Galactic plane spanning from $l=24^\circ$ to $l=30^\circ$ \citep{F06, D07, C09}. The line of sight to these clusters crosses several Galactic arms and reaches an area of very high obscuration at roughly the same distance as they do \citep{NGF10}. They also lie amidst an area of high stellar density. Under these conditions, the RSGs are the only discernible population. Using population synthesis models it has been inferred that, to host so many RSGs, the clusters must be very massive \citep[eg.][]{D07, C09b}.
   
   The first cluster found, RSGC1 \citep{F06}, is the innermost, youngest, and most obscured, with an estimated age of $\tau=12\pm2~\mathrm{Myr}$ and $M_{\mathrm{initial}}=3\pm1\times 10^{4}~M_{\odot}$. Alicante 8, located in its inmediate vicinity, seems slightly older and somewhat smaller \citep{NGF10}. Moving outwards from the Galactic centre, the next cluster to be found is RSGC2=Stephenson 2; it is the least obscured and most massive, with $\tau=17\pm3~\mathrm{Myr}$ and $M_{\mathrm{initial}}=4\pm1\times10^4~$ \citep{D07}.
   
   RSGC3 is the most isolated of the clusters, around $l=29^\circ$. It has an estimated $\tau=16-20~\mathrm{Myr}$ and $M_{\mathrm{initial}}=2-4\times10^4~M_{\odot}$ \citep{C09, A09}. More interestingly, an extended association of RSGs has been detected around it, harbouring at least another smaller cluster, Alicante 7 \citep[][hereafter N11]{NGF11}. Photometric studies suggest that the RSGC3 complex contains at least 50 RSGs, pointing towards an initial mass in excess of $60~000~M_{\odot}$, and quite possibly around $100~000~M_{\odot}$ (N11).
   
   To confirm these numbers, it has been shown that medium-resolution spectroscopy is needed. The low temperatures and surface gravities present in RSGs make for a spectrum with very prominent molecular bands, allowing a quick and easy assessment of their nature. Also, with an uncertainty in the velocity along the line of sight ($v_{\mathrm{los}}$)  of $5~\mathrm{\mathrm{kms^{-1}}}$ it is possible to separate the RSGs pertaining to the foreground Galactic arms from those related to RSGC3 (N11). The two favoured wavelength ranges for this have been the area around the calcium triplet at $8500~\AA{}$ (N11) and the near infrared CO feature at $2.3~\mathrm{\mu m}$ \citep{D07}. The former has the advantage of being a much more studied region, where a separation of spectral types is easier, while the latter profits from the intrinsic high luminosity of the sources under study and the higher transparency of the interstellar medium at these wavelengths.
   
   In this work we complement the pilot programme presented in N11 using near-infrared spectroscopy to study some of the more obscured sources around RSGC3.

%__________________________________________________________________

\section{Target selection and observations}

\begin{figure*}
\centering
\includegraphics[width=12cm, angle=90]{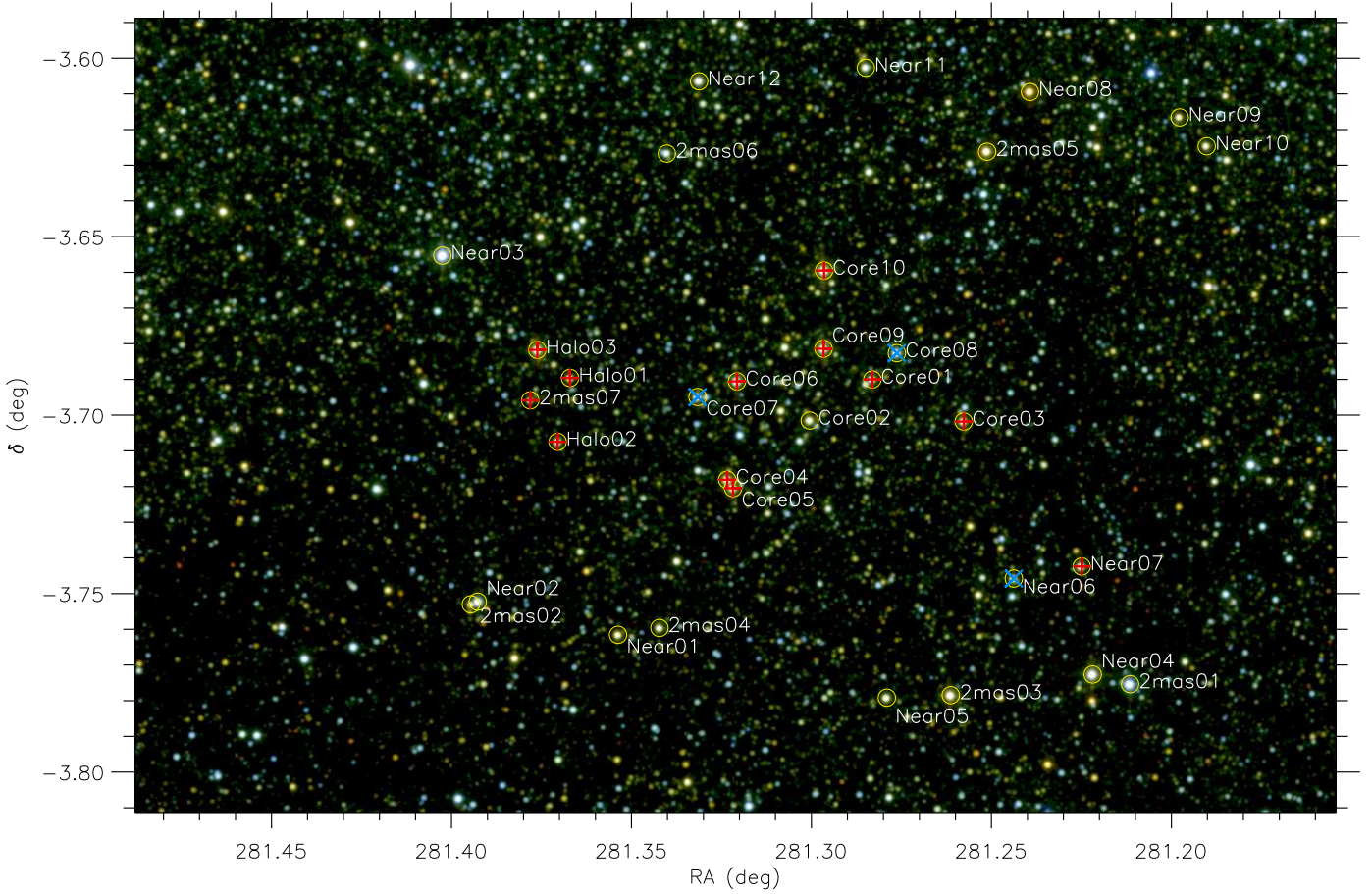}
\caption{False colour image from 2MASS (blue channel for $J$ band, green for $H$, and red for $K_\mathrm{S}$), covering approx. 13'x20', with the sources under study marked. Labels follow the notation in Table \ref{tab1}. Yellow circles denote the photometric candidates, while spectroscopically confirmed RSGs are marked with a red plus. Blue crosses left for those stars that are late type giants. North is up, east to the left.}
\label{rgb}
\end{figure*}

RSGC3 appears as a concentration of very bright infrared sources \citep{C09, A09}, but surrounding it there is a more diffuse population of RSGs already pointed out by \citet{C09}, although their study was constrained to a 7' field. This extended population was addressed, now over a much wider area with a 18' radius from the cluster centre, in N11. Although the traditional photometric criteria used to pick RSGs fail in very reddened, crowded fields, it is possible to overcome this difficulty using the known properties of the already confirmed members of RSGC3 and infrared photometry (N11).

\subsection{Photometric criteria}
\label{crit}
Using the photometry from 2MASS \citep{S06}, it is possible to define a reddening-free parameter $Q=(J-H)-1.8\times(H-K_\mathrm{S})$ that allows the separation of early type stars (with $Q\lesssim0.0$) from late-type main sequence stars and red clump giants (with $Q\gtrsim0.4$). RSGs often deredden to the position of yellow stars ($Q\approx 0.2-0.3$), though a non-negligible fraction reach values of $Q\approx0.4$ \citep{NGF10,Me11}.

For this study the selection criteria are the same as those used in N11. The candidate RSGs must comply with these conditions:
\begin{enumerate}
\item The reddening-free parameter $Q$ should fall between those of red and blue stars: $0.1\le Q \le0.4$.
\item The RSGs surrounding RSGC3 are intrinsically red and expected to be affected by interstellar extinction, so they should have $(J-K_\mathrm{S})>1.3$.
\item The typical magnitudes for the confirmed members of RSGC3 are in the $K_\mathrm{S}=5-6$ range, so we choose a cutoff at $K_\mathrm{S}\le7.5$ that takes the possible effects of differential extinction in the field into account.
\item As RSGs have high visible to infrared colours, we impose an additional cut at $i>9.5$ using DENIS \citep{E97} when available (since the whole field around RSGC3 is not covered) and USNO-B1.0 elsewhere. The candidates were also checked against DSS2, because they must be invisible in its blue plates.
\end{enumerate}
While in N11 the authors also had the constraint that, to be observable, the sources needed to have $i<15.5$, as this study will be carried out in the infrared, we are not affected by it.

There are $\sim50$ candidates in an area with radius in the range $7'\le r\le20'$ around RSGC3. The distribution of these stars is not homogeneous, and some overdensities are evident. While most of them have already been covered in N11, some 16' southwards of RSGC3 there is a clear group of some 20 candidates. As can be seen in Fig. \ref{rgb}, these stars form a main overdensity composed of ten candidates (labelled {\em Core}), flanked by a smaller, isolated blob of three more candidates, $30"$ to the east (labelled {\em Halo}). Surrounding these two groups there is a more diffuse population of another ten possible RSGs ({\em Near} sources). The properties of these objects are listed in Table \ref{tab1}.

\begin{table*}
\caption{\label{tab1}Photometrically selected candidates and complementary targets, along with their characteristics}
\centering
\begin{tabular}{@{}cccrrrccccc@{}} 
\hline\hline
ID&2MASS&$i$\tablefootmark{1} & $J$\tablefootmark{1} & $H$\tablefootmark{1} & $K_\mathrm{S}$\tablefootmark{1} & Q & $E(J-K_\mathrm{S})$ & $v_{\mathrm{LSR}}$  & $EW_{\mathrm{CO}}$& Spectral\\
 & & & & & & & &($\mathrm{kms^{-1}}$) &  (\AA{}) & type\\
\hline 
Core01&18450794$-$0341240&    ...    &10.50$\pm$0.03&8.08$\pm$0.04&6.91$\pm$0.02&0.31&2.60& 91&$-$22.29&K7I\\
Core02&18451210$-$0342054&    ...    &10.42$\pm$0.02&8.04$\pm$0.02&6.90$\pm$0.02&0.33& ...&  ... & ... &...\\
Core03&18450185$-$0342065&16.90$\pm$0.13&10.45$\pm$0.02&7.94$\pm$0.03&6.70$\pm$0.01&0.29&2.69& 95&$-$27.08&M1I\\
Core04&18451760$-$0343051&    ...    &9.97$\pm$0.02&7.40$\pm$0.03&6.06$\pm$0.02&0.15&2.81& 95&$-$27.69&M2I\\
Core05&18451722$-$0343136&    ...    &10.38$\pm$0.02&7.64$\pm$0.03&6.28$\pm$0.02&0.29&3.04& 95&$-$26.71&M1I\\
Core06&18451696$-$0341260&    ...    &9.56$\pm$0.02&7.29$\pm$0.04&6.15$\pm$0.02&0.22&2.23& 91&$-$30.33&M4I\\
Core07&18451959$-$0341417&    ...    &10.00$\pm$0.02&7.92$\pm$0.03&6.85$\pm$0.02&0.15&2.00&134&$-$20.70&M3III\\
Core08&18450629$-$0340575&11.4$\pm$0.3\tablefootmark{a}&8.60$\pm$0.02&7.33$\pm$0.03&6.84$\pm$0.02&0.37&0.62& 50&$-$20.40&M3III\\
Core09&18451118$-$0340531&15.4$\pm$0.3\tablefootmark{a}&9.34$\pm$0.02&6.86$\pm$0.04&5.62$\pm$0.02&0.23&2.62& 86&$-$27.93&M2I\\
Core10&18451116$-$0339341&14.8$\pm$0.3\tablefootmark{a}&8.80$\pm$0.02&6.47$\pm$0.03&5.35$\pm$0.02&0.29&2.39& 95&$-$26.85&M1I\\
Halo01&18452809$-$0341225&    ...    &11.49$\pm$0.02&8.73$\pm$0.02&7.31$\pm$0.02&0.21&3.16&119&$-$25.63&M0I\\
Halo02&18452892$-$0342269&    ...    &10.75$\pm$0.02&8.26$\pm$0.03&7.07$\pm$0.02&0.33&2.66& 42&$-$24.59&M0I\\
Halo03&18453027$-$0340542&    ...    &9.21$\pm$0.02&6.84$\pm$0.03&5.64$\pm$0.02&0.21&2.56& 85&$-$24.04&K8I\\
Near01&18452489$-$0345415&    ...    &11.34$\pm$0.02&8.47$\pm$0.03&7.04$\pm$0.01&0.30& ...&  ... & ... &...\\
Near02&18453424$-$0345081&12.2$\pm$0.3\tablefootmark{a}&8.73$\pm$0.02&7.23$\pm$0.03&6.57$\pm$0.03&0.31& ...&  ... & ... &...\\
Near03&18453660$-$0339191&10.42$\pm$0.03&7.19$\pm$0.02&5.82$\pm$0.03&5.23$\pm$0.01&0.30& ...&  ... & ... &...\\
Near04&18445323$-$0346216&13.96$\pm$0.03&8.51$\pm$0.02&6.53$\pm$0.02&5.63$\pm$0.01&0.35& ...&  ... & ... &...\\
Near05&18450697$-$0346449&    ...    &10.40$\pm$0.02&8.08$\pm$0.02&6.94$\pm$0.02&0.29& ...&  ... & ... &...\\
Near06&18445850$-$0344448&    ...    &10.66$\pm$0.02&8.45$\pm$0.03&7.33$\pm$0.03&0.18&2.19&114&$-$20.27&M3III\\
Near07&18445398$-$0344326&16.88$\pm$0.12&10.16$\pm$0.02&7.89$\pm$0.03&6.76$\pm$0.01&0.25&2.26&104&$-$29.44&M3I\\
Near08&18445742$-$0336341&    ...    &10.92$\pm$0.02&7.53$\pm$0.03&5.80$\pm$0.02&0.29& ...&  ... & ... &...\\
Near09&18444745$-$0336596&    ...    &11.23$\pm$0.02&8.58$\pm$0.02&7.30$\pm$0.02&0.35& ...&  ... & ... &...\\
Near10&18444564$-$0337288&    ...    &10.96$\pm$0.02&8.44$\pm$0.02&7.20$\pm$0.01&0.28& ...&  ... & ... &...\\
Near11&18450837$-$0336096&    ...    &9.98$\pm$0.02&7.97$\pm$0.02&6.98$\pm$0.02&0.23& ...&  ... & ... &...\\
Near12&18451949$-$0336233&    ...    &9.56$\pm$0.02&7.54$\pm$0.04&6.50$\pm$0.01&0.16& ...&  ... & ... &...\\
\hline
\multicolumn{11}{c}{Complementary targets}\\
\hline
2mass01&18445076$-$0346312&9.96$\pm$0.03&7.61$\pm$0.01&6.43$\pm$0.02&6.00$\pm$0.02&0.41& ...&  ... & ... &...\\
2mass02&18453470$-$0345107&    ...    &12.26$\pm$0.04&9.19$\pm$0.03&7.61$\pm$0.02&0.21& ...&  ... & ... &...\\
2mass03&18450271$-$0346424&16.96$\pm$0.13&9.61$\pm$0.02&6.99$\pm$0.02&5.79$\pm$0.02&0.46& ...&  ... & ... &...\\
2mass04&18452213$-$0345345&17.2$\pm$0.3\tablefootmark{a}&10.92$\pm$0.02&8.46$\pm$0.03&7.34$\pm$0.02&0.43& ...&  ... & ... &...\\
2mass05&18450029$-$0337344&    ...    &9.97$\pm$0.02&7.19$\pm$0.02&5.89$\pm$0.02&0.43& ...&  ... & ... &...\\
2mass06&18452164$-$0337363&13.9$\pm$0.3\tablefootmark{a}&10.24$\pm$0.03&8.47\tablefootmark{b}&7.69\tablefootmark{b}& 0.36& ...&  ... & ... &...\\
2mass07&18453075$-$0341451&    ...    &11.81$\pm$0.02&9.57$\pm$0.02&8.51$\pm$0.02&0.34&2.28& 66&-24.82&M0I\\

\hline
\end{tabular}
\tablefoot{
\tablefoottext{1}{$JHK_\mathrm{S}$ magnitudes come from 2MASS, $i$ magnitudes from DENIS except where otherwise noted.}
\tablefoottext{a}{Photographic $I$ magnitude from USNO.}
\tablefoottext{b}{The magnitude is an upper limit.}
}
\end{table*}

\subsection{Spectroscopy}
\label{secspec}
Observations were carried out using the VLT (UT3) and ISAAC at Cerro Paranal (Chile), during a run of one night on  2011 July 9. Since velocities along the line of sight are required and the targets are bright, we opted for medium resolution and a 0.3" slit, so as to maximize resolving power. Using a central wavelength of $2.27~\mathrm{\mu m}$, we were able to cover the CO bandhead at $2.3~\mu\mathrm{m}$ while avoiding any major telluric absorption. With this setup, the mean measured resolving power is $R=9000$. 

Although ISAAC is a long-slit spectrograph, it is easy to align two objects over the slit, attaining some multiplexing. To profit from this, we paired our targets according to brightness and distance, so as to always have two targets per pointing. If any object was left out, a small set of fillers was drawn from those stars that in the field that failed one of the selection criteria listed in Sect.~\ref{crit}. These sources are listed as complementary targets in Table \ref{tab1}. The programme was complemented with radial velocity standards and some targets from Stephenson 2 (N11). As the stars from this cluster have been observed in the calcium triplet region, they could be used as a test for the infrared spectral classification used here.

Unluckily, the weather was very unstable during the night, resulting in a very variable seeing and time losses due to high wind and clouds. Due to this, only $\sim50\%$ of the programme was completed; beyond the stars detailed in Table \ref{tab1}, two stars from Stephenson 2 and two velocity standards were observed. Thanks to the intrinsic brightness of the targets, even under these conditions, very high signal-to-noise ratios can be obtained with short exposure times. The usual ABBA scheme was applied, and with integrations of $15~\mathrm{s}$ we guaranteed a nominal $S/N\geq500$ for all the target stars.

Reduction was performed using {\em IRAF} \citep{iraf} standard routines. Flats and arc frames were obtained for each target, and telluric absorption was corrected using an early-type (usually around B9) star observed at roughly the same airmass as the object. These two calibrations are the main source of error in the whole reduction process, so it is important to check their performance before analysing the spectra. We perform these checks using these early-type stars. In the wavelength range observed, these do not present any significant absorption line, and so all the features in a spectrum are telluric effects. We can use these lines to check the stability and accuracy of the wavelength solution adopted, comparing the measured centres of the most significant lines with the same absorptions measured over a telluric absorption template. There is a systematic difference of $-1.1~\mathrm{km\;s^{-1}}$, with a standard deviation of $1.4~\mathrm{km\;s^{-1}}$. While the systematic will be corrected in a later step, this last figure gives us our internal error in velocity. To check the telluric absorption correction, we corrected all the standard stars using their closest match in airmass, and we compared the resultant spectra with a synthetic template. The typical difference is around $2\%$, although for narrow areas associated with molecular absorption it can reach a maximum of $8\%$.

\section{Spectral classification}
\subsection{Spectral types and luminosities}
\begin{figure}
\resizebox{\hsize}{!}{\includegraphics{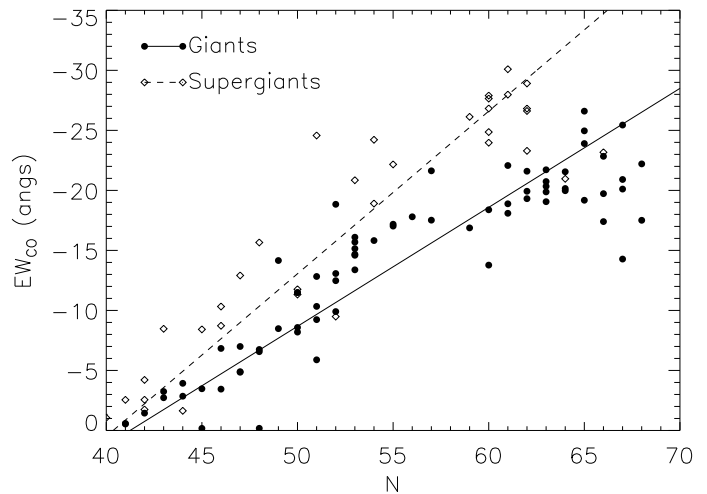}}
\caption{Calibration of the spectral type, denoted by the numeral $N$ (see text) and the equivalent width of the CO bandhead, $EW_{\mathrm{CO}}$, for giants (filled circles and continuous line) and supergiants (diamonds and dashed line).}
\label{CON}
\end{figure}

As discussed in \citet{D07} and \citet{CGF08}, the equivalent width of the  CO bandhead at $2.3~\mu \mathrm{m}$ is a good indicator of temperature, and ir can be used for spectral classification. To obtain this equivalent width, we use the wavelength ranges described in Table \ref{coreg}, and the calibration between $EW_{\mathrm{CO}}$ and spectral type from \citet{NGF10}.

\begin{table}
\caption{\label{coreg}Wavelength ranges used for the derivation of $EW_{\mathrm{CO}}$, from \citet{CGF08}}
\centering
\begin{tabular}{cc} 
\hline\hline
Region&Range \\
&(\AA{})\\
\hline
$^{12}\mathrm{CO}$ Absorption&$22910<\lambda\leq23020$\\
Continuum \#1&$22300<\lambda\leq22370$\\
Continuum \#2&$22420<\lambda\leq22480$\\
Continuum \#3&$22680<\lambda\leq22790$\\
Continuum \#4&$22840<\lambda\leq22910$\\
\hline
\end{tabular}
\end{table}

In that work the authors relate those quantities using the library of \citet{R09}. As can be seen in Fig. \ref{CON}, if we denote the spectral type by a numeral $N$ ($N=40$ for a G0 star, $N=50$ for K0, $N=59$ for M0 and so forth), this quantity is a linear function of $EW_{\mathrm{CO}}$ in the range $40<N\lesssim64$ (between G0 and M5), both for giants and supergiants:
\begin{eqnarray*}
N(\mathrm{G})=41.24-1.01\times EW_{\mathrm{CO}}\\
N(\mathrm{SG})=40.36-0.74\times EW_{\mathrm{CO}}.
\end{eqnarray*}

According to \citet{D07} the typical error of this calibration is $\pm2$ subtypes, although our very limited sample of stars from Stephenson 2, whose spectral types have been measured in the Ca triplet region, points towards smaller deviations (Table \ref{ste2}). In this error budget the uncertainties in the calculation of  $EW_{\mathrm{CO}}$ should be included. According to \citet{ewerr}, the error in the derivation of an equivalent width can be approximated by
 \begin{equation}
 \sigma(EW)=\sqrt{2M}\frac{\bar{F}}{\sigma(F)}
 \end{equation}
 where $M$ is the number of pixels used in the calculation, $\bar{F}$ the mean flux measured in them, and $\sigma(F)$ the noise associated with this measurement (shot and readout noises, etc.). This, for all our spectra, gives an upper limit for the error of $0.2~$\AA{}. We can readily see that this has little significance in calculating the spectral type (less than $0.2$ spectral types), and the largest source of error will be that induced by the transformation between $N$ and $EW_{\mathrm{CO}}$. Due to the limited comparison sample we have for estimating this uncertainty, we use the more conservative error estimation from \citet{D07}.

\begin{table}
\caption{\label{ste2}Measured $EW_{\mathrm{CO}}$ of stars from Stephenson 2 and derived spectral types}
\centering
\begin{tabular}{cccc} 
\hline\hline
Star\tablefootmark{a}&&Spectral type&\\
&This work&D07\tablefoottext{1}&N11\tablefoottext{2}\\
\hline
D18&M1I&M4I&M1.5Iab\\
D20&M0I&M2I&M1.5Iab\\
\hline
\end{tabular}
\tablefoot{
\tablefoottext{a}{The nomenclature is taken from \citet{D07}.}\newline
\tablefoottext{1}{\citet{D07}, based on their own $EW_{\mathrm{CO}}$ calibration.}\newline
\tablefoottext{2}{N11, based on Ca triplet/TiO indexes.}
}
\end{table}

Although clear luminosity indicators in the infrared have not been yet established for this kind of star, it is possible to distinguish between giants and supergiants on the basis of their $EW_{\mathrm{CO}}$. As explained in \citet{D07}, any star with $EW_{\mathrm{CO}}\le -24~\AA{}$ is clearly a class I star, while stars with $-22\ge EW_{\mathrm{CO}}\ge-24~\AA{}$ are very likely supergiants too, although there are some interlopers of class II/III in this region. Finally, all giants have $EW_{\mathrm{CO}}\ge -22~\AA{}$.

Once a spectral type has been assigned to a source based on its $EW_{\mathrm{CO}}$, since all these sources have 2MASS photometry, it is possible to calculate the colour excess $E(J-K_\mathrm{S})$, which gives us a measure of the interstellar extinction affecting the targets. For this we use the temperature-colour calibration of \citet{L05}. Once a spectral type is known, a temperature can be assigned to the star, and from it a value of  $(J-K_\mathrm{S})_0$ derived. Because giants and supergiants have very similar broad-band colours, we also calculate $(J-H)_0$ and $(H-K_\mathrm{S})_0$ using the tabulated values for giants from \citet{Stra09}, interpolating when necessary. We find both studies to be in very good agreement, with differences in $(J-K_\mathrm{S})_0$ below $0.02$ magnitudes. For a detailed account of these values, see Appendix \ref{JK0}. The typical change in colour between consecutive spectral types is $\Delta(J-K_\mathrm{S})_0\sim0.07$ for late stars, and so accounting for the $\pm2$ spectral types uncertainty associated with the $EW_{\mathrm{CO}}$ calibration implies a $\pm0.15$ error in the derived $E(J-K_\mathrm{S})$.

\subsection{Radial velocities}

Cross correlation against some model spectrum or well-calibrated observation is the most effective way to derive the Doppler shift associated with the displacement of the source along the line of sight. With a proper selection of this spectral template, and assuming that the stellar spectrum under study has  enough lines, accuracies as high as $\lesssim10\%$ of the spectral resolution can be achieved.

Because all late type stars present rather similar spectra around $2.3~\mu \mathrm{m}$, dominated by molecular absorptions, there is no need to use different templates (as in N11), so we opt to follow the procedure outlined in \citet{D07}, using the spectrum of Arcturus, from \citet{WH96}, for comparison. Derived $v_{\mathrm{los}}$ with this method were transformed into the local standard of rest (LSR) reference system using {\em IRAF}s {\em rvcorrect} package, using the standard solar motion ($+20~\mathrm{km\;s^{-1}}$ towards $l=56^\circ,~b=23^\circ)$. 

As shown is Sect. \ref{secspec}, there is a small systematic residual in our wavelength calibration. This effect adds up with the wavelength displacement of our template spectrum: the data from \citet{WH96} have not been transformed to any standard frame of reference (LSR or heliocentric). This, along with the Doppler shift of Arcturus, measured at $-5.2~\mathrm{km\;s^{-1}}$ \citep{arcturus} and its pulsation, introduces an offset in all the velocities derived with this method. We estimate this offset by comparing our results for the velocity standards and the stars from Stephenson 2, as can be seen in Table \ref{rvel}. From these values we obtain a mean shift of $-15.1~\mathrm{km\;s^{-1}}$ and a dispersion of $3~\mathrm{km\;s^{-1}}$. This later amount also serves as the net error in our derived velocities.

\begin{table}
\caption{\label{rvel} Uncorrected $v_{\mathrm{los}}$ of the observed stars from Stephenson 2 and the velocity standards.}
\centering
\begin{tabular}{ccc} 
\hline\hline
Star\tablefootmark{a}&\multicolumn{2}{c}{$v_{\mathrm{los}}$}\\
&\multicolumn{2}{c}{$\mathrm{(km~s^{-1})}$}\\
&This work&Literature\\
\hline
D18\tablefootmark{1}&93.65&111.20\\
D20\tablefootmark{1}&93.65&109.90\\
HD125184\tablefootmark{2}&$-$28.14&$-$12.44\\
HD190007\tablefootmark{2}&$-$42.04&$-$31.03\\
\hline
\end{tabular}
\tablefoot{
\tablefoottext{a}{The nomenclature is taken from \citet{D07}.}\newline
\tablefoottext{1}{Velocities with respect to the local standard of rest, taken from \cite{D07}.}\newline
\tablefoottext{2}{Heliocentric velocities, taken from \citet{W07}.}
}
\end{table}

\section{Results}
\subsection{Confirmed supergiants}

\begin{figure}
\resizebox{\hsize}{!}{\includegraphics[angle=90]{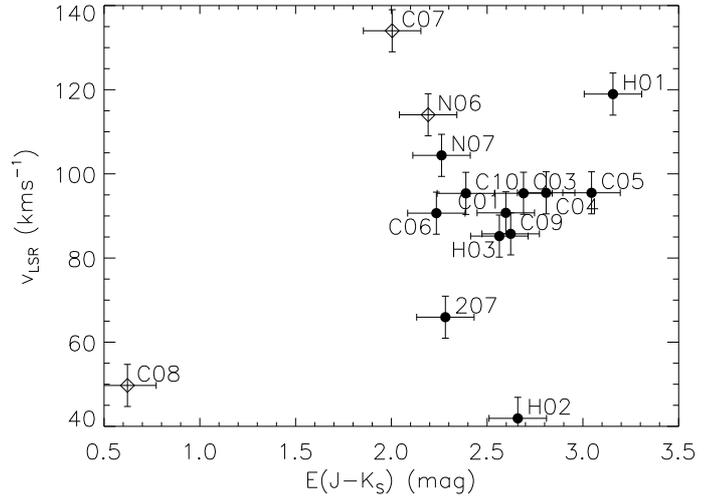}}
\caption{Colour excess versus $v_{\mathrm{LSR}}$ for the observed sources. Filled circles denote supergiants, while empty diamonds are for giants. The notation is shorthand for the identificators in Table \ref{tab1} being C for Core, H for Halo, N for Near and 2 for complementary targets. There is a clear overdensity of RSGs around $E(J-K_\mathrm{S})=2.6$ and $v_{\mathrm{LSR}}=92~\mathrm{km~s^{-1}}$.}
\label{vlsrvsejk}
\end{figure}

All the photometrically selected targets have $EW_{\mathrm{CO}}<-20~\AA{}$ (Table \ref{tab1}). This implies that they are luminous, intrinsically red objects, with types above M3 or K6 depending on whether they are giants or supergiants.  Of the 15 observed candidates, 11 turned to have $EW_{\mathrm{CO}}<-24$, so they are bona fide supergiants.  Another object, Core01, is very likely a member of this class, with $EW_{\mathrm{CO}}=-22.3~\AA{}$. Although with this equivalent width, this star could be both an M giant or a K supergiant, we opt for the latter classification since it has a $Q$ value consequent with an RSG, plus its $E(J-K_\mathrm{S})$ and $v_{\mathrm{LSR}}$ are very similar to the main body of supergiants (see Table \ref{tab1} and Fig. \ref{vlsrvsejk}). This suggests that all the sources lie at roughly the same distance and under the same reddening, so Core01 should then have a similar absolute magnitude to the other RSGs, which is too bright to be a giant.

As can be seen in Fig. \ref{vlsrvsejk}, eight of the RSGs fall in a relatively narrow range of  colour excess and velocity, with $85<v_{\mathrm{LSR}}<95~\mathrm{kms^{-1}}$ and $2.2<E(J-K_\mathrm{S})<3.0$. This points towards being physically associated, hence forming a new cluster, which we identify as Alicante 10. Although the velocity dispersion might seem high, supergiant clusters are known to show these values (N11). In part, this is due to the intrinsic pulsation of the stars; in \citet{Me08} the authors perform several $v_{\mathrm{los}}$ measurements for various red supergiants, finding dispersions as high as $5~\mathrm{km\;s^{-1}}$. This intrinsic variation introduces an extra, non-negligible term to the velocity dispersion of the cluster. 

The remaining supergiants, although under a similar amount of extinction, have very different $v_{\mathrm{LSR}}$, ranging from $42~\mathrm{kms^{-1}}$ to $119~\mathrm{kms^{-1}}$. Both these velocities and the presence of a cluster will be explored in subsequent sections.

\begin{figure*}
\resizebox{\hsize}{!}{\includegraphics{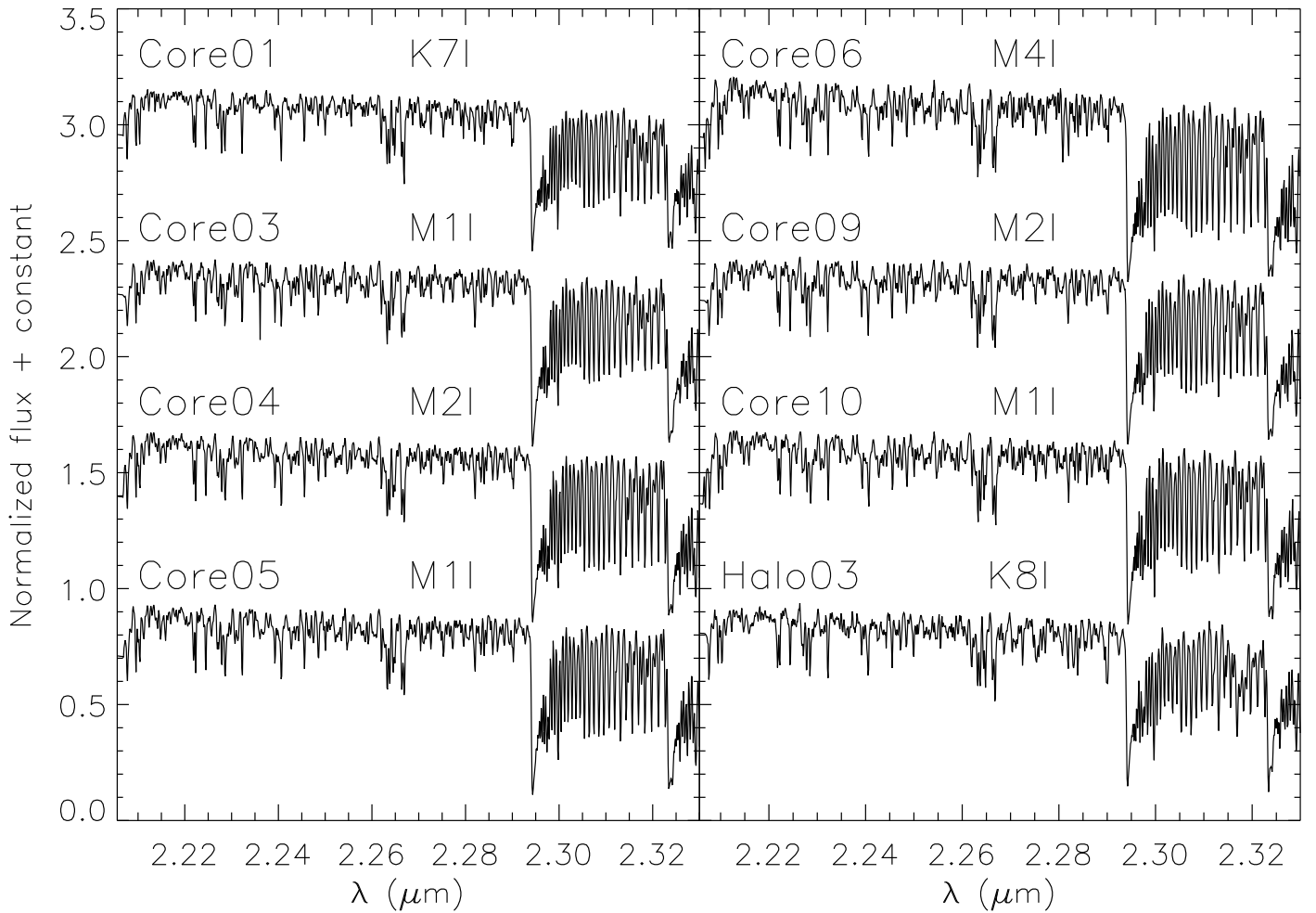}}
\caption{K band spectra of the eight proposed members of Alicante 10.}
\label{specs}
\end{figure*}

\subsection{The new cluster Alicante 10}

\begin{figure}
\resizebox{\hsize}{!}{\includegraphics[angle=90]{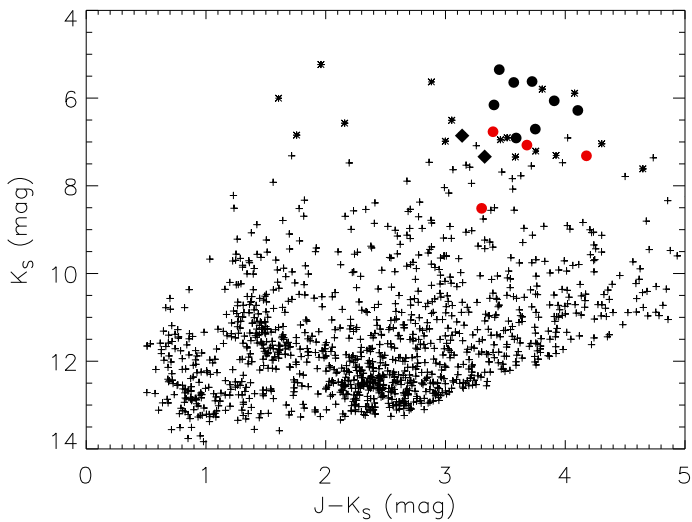}}
\caption{Colour magnitude diagram of all the 2MASS sources around 10' from the centre of Alicante 10. Spectroscopically observed sources are marked with full dots for confirmed RSGs, (black for Alicante 10 members, red for other sources) and diamonds for giants (class III);  stars mark photometrically selected candidates whose stellar nature has not been confirmed spectroscopically.}
\label{dcm}
\end{figure}

The new cluster, Alicante 10 --as for many other RSG clusters-- appears to be a rather diffuse blob at bright $K_\mathrm{S}$ magnitudes in the redder half of a $J-K_\mathrm{S}$ vs. $K_\mathrm{S}$ diagram (Fig. \ref{dcm}). Based solely on this diagram, it could be possible to assign some five other sources to the cluster (contained in Table \ref{tab1}), but since it is very hard to differentiate between RSGs and late giants only on photometric grounds, we opted to restrict ourselves to the spectroscopically confirmed membership. This implies that Alicante 10 has a population of some eight RSGs. If the cluster has the same age as its neighbour RSGC3, it would mean an initial mass well in excess of $10~000~M_{\odot}$ and approaching $20~000~M_{\odot}$ \citep{C09b}, although this number decreases significantly with age.

\begin{figure}
\resizebox{\hsize}{!}{\includegraphics{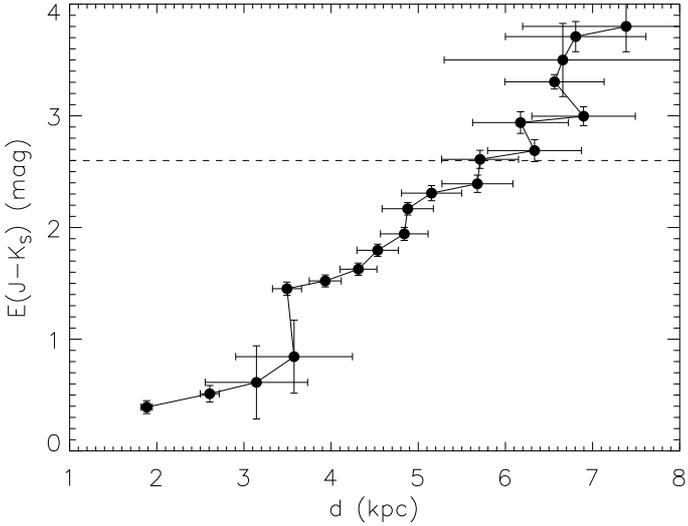}}
\caption{Colour excess as a function of distance, derived using the red clump giants and the methods of \citet{CCGF08}. The striped line marks the mean colour excess of the cluster.}
\label{dvsejk}
\end{figure}

To estimate the distance to the cluster, we rely on the $d$ vs. $E(J-K_\mathrm{S})$ diagram. Using all the red clump giants in an circular area with a radius of 10' around the cluster and the method described in \citet{CCGF08}, it is possible to sample the extinction along the line of sight. Since this is an in-plane field (hence subject to high reddening) to reach the red clump giants at the same distance as the RSGs of Alicante 10, we need to use GPS-UKIDSS photometry \citep{WC07}. The results can be seen in Fig. \ref{dvsejk}. After a sudden increase at $\sim3.5~\mathrm{kpc}$, the evolution of $E(J-K_\mathrm{S})$ with distance is rather smooth, although its gradient seems to grow with $d$. The red clump giants reach the mean $E(J-K_\mathrm{S})$ of the RSGs of Alicante 10 around $6~\mathrm{kpc}$. This estimate is in concordance with the other clusters in the area (N11).

\section{Discussion}
\subsection{Alicante 10 and its context: The RSGC3 complex}
The RSGC3 complex (containing RSGC3, Alicante 7 and Alicante 10) is an area that is particularly rich in supergiants. At least 35 have already been spectroscopically confirmed \citep[][N11 and this work]{C09}, although the number is surely much higher, well above 50. Such a concentration of RSGs would imply a very high initial mass, well in excess of $100~000~M_\odot$, but whether these stars are indeed an association or just an effect of projection (since the line of sight is tangent to the Scutum/Crux arm or the molecular ring) remains an open question.

We can compare the estimate of the distance based on $E(J-K_\mathrm{S})$ with another one based on the rotation curve of the Galaxy, making use of the measurements of $v_{\mathrm{LSR}}$. If one assumes a rotation model for the Galaxy, knowing the velocity along the line of sight of a source and its coordinates, it is possible to invert its distance, although degenerate into two values. Because the $v_{\mathrm{los}}$ vs. $d$ is symmetric around its maximum, one usually obtains a near estimate, where the $v_{\mathrm{los}}$ of the sources is reached before the galactocentric tangent point (the maximum of $v_{\mathrm{los}}$ in that line of sight) and another far distance value where the same velocity occurs beyond the tangent point. To break this degeneracy, we use the extinction corrected magnitude $K_\mathrm{e}$. This value only depends on the intrinsic magnitude of the source and its distance. Since we know the colour excess of our stars, we can derive
\begin{equation}
K_\mathrm{e}=K_\mathrm{S}-\frac{E(J-K_\mathrm{S})}{A_\mathrm{J}/A_\mathrm{K}-1}
\end{equation}
where $E(J-K_\mathrm{S})$ comes from Table \ref{tab1} and $A_\mathrm{J}/A_\mathrm{K}$ from \citet{RL85}.

\begin{figure}
\resizebox{\hsize}{!}{\includegraphics[angle=90]{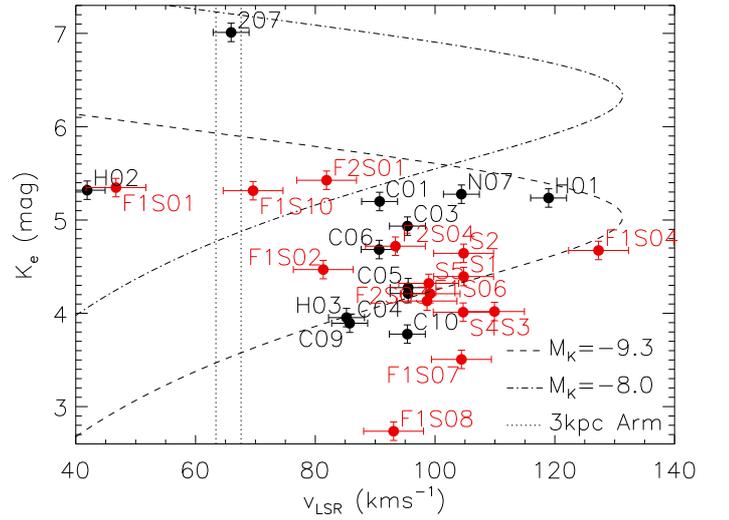}}
\caption{$v_{\mathrm{LSR}}$ versus extinction-corrected magnitude $K_\mathrm{e}$ for the observed supergiants. Black dots denote the measurements derived in this work, while red dots are used for those taken from N11. The dashed and dot-dashed lines represent the locus that a supergiant population with mean $M_\mathrm{K}=-9.3$ (that corresponds to the bulk of RSGC3) and $M_\mathrm{K}=-8.0$ \citep[all RSGs are expected to be brighter than this limit,][]{W83} would occupy assuming the rotation curve from \citet{Re09}. This work uses $\Theta_{0}=254~\mathrm{km\;s^{-1}}$ and $R_0=8.4~\mathrm{kpc}$. The dotted lines mark the expected velocities of the $3~\mathrm{kpc}$ arm at $l=28^\circ$ and $l=29^\circ$, according to \citet{Da08}}
\label{vlsrvske}
\end{figure}

We can compare the measured $(v_{\mathrm{LSR}},K_\mathrm{e})$ with different analytic curves using several values of $M_\mathrm{K}$, once we assume a model for the rotation curve of the Milky Way \citep[we use][]{Re09}. Of these curves, we choose those that fit best the gross of the stars from Alicante 10 and RSGC3. The data for this last cluster are taken from N11, with an updated estimation of the $v_{\mathrm{LSR}}$ (for homogeneity with this work) according to the method described in Negueruela et al. (in prep.). We retain the nomenclature of the authors, since they break down their sources into three fields, one covering RSGC3 (stars named SN, with N from 1 to 5) and two other mapping adjacent locations where other surrogate clusters might be located, Fields 1 and 2 (with notations F1SNN and F2SNN). There is a fourth field in their study (F3SNN), but since it does not contain any spectroscopically confirmed RSG, it has been left out. Apart from RSGC3, the stars in Field 1 are of particular interest, as it is there where the authors find the new cluster Alicante 7. The results of these calculations are displayed in Fig. \ref{vlsrvske}.

As can be seen, once we account for the effect of extinction, the stars from Alicante 10 and RSGC3 have similar infrared brightness, although Alicante 10 seems to have a wider range of variation, particularly towards the faint end. This could be, in part, because the spectra from N11 are gathered in the I band, and this could bias the study against the fainter component of RSGC3, which would remain invisible outside the near infrared.

As shown in the previous section, the bulk of Alicante 10 has a velocity $v_{\mathrm{LSR}}=92~\mathrm{km\;s^{-1}}$, while in Fig. \ref{vlsrvske} the stars from RSGC3 appear concentrated at $v_{\mathrm{LSR}}=105~\mathrm{km\;s^{-1}}$. This translates into $d=5.9\pm0.3~\mathrm{kpc}$ for the latter and $d=5.1\pm0.2~\mathrm{kpc}$ for the former. Although the estimation for RSGC3 agrees nicely with the one based on the red clump method (based on its mean $E(J-K_\mathrm{S})$ the derived distance in N11 is $d\sim6~\mathrm{kpc}$), this is not the case for Alicante 10.There are several factors that need to be accounted before drawing any conclusion from this difference.

There are several possible sources of error that need to be examined. Although the derivation of the colour excess depends on a correct estimation of the spectral type, we have already shown that this translates into an uncertainty of $\pm0.15$ magnitudes. The luminosity function of the red clump is well calibrated, but it has a minor dependency with metallicity \citep{CCGF08}. This could result in an extra $\pm0.1$ on $E(J-K_\mathrm{S})$. Adding all of these to the uncertainties associated to the transformation from $K_\mathrm{S}$ to $d$ for the red clump giants one could push the distance estimation $0.5~\mathrm{kpc}$ away \citep[the typical error found in ][]{CCGF08}, not enough to bring it in agreement with the dynamical estimation.

\begin{figure}
\resizebox{\hsize}{!}{\includegraphics[angle=90]{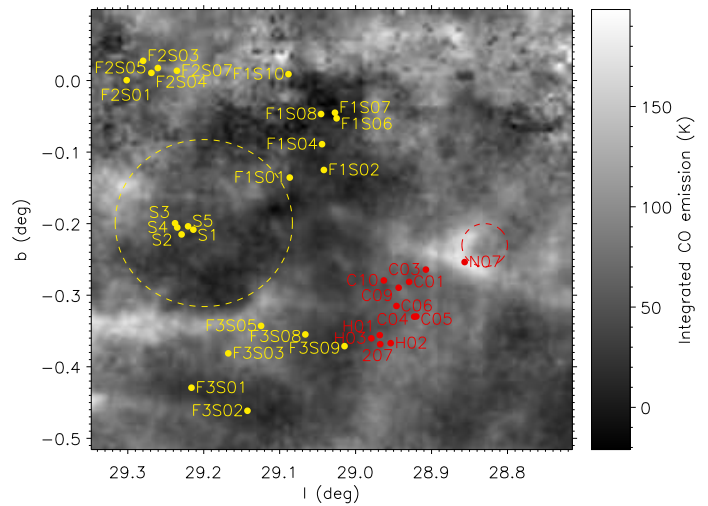}}
\caption{Spatial distribution of the sources from this work (red dots) and N11 (yellow dots), plotted over the CO emission (integrated in the range $0<v_{\mathrm{LSR}}<130~\mathrm{km~s^{-1}}$, the tangent point along the line of sight) taken from the Galactic Ring Survey \citep{putaringonit}. The yellow dashed circle represents the position of RSGC3 (assuming a radius for the cluster of 7'), and the red dashed circle follows the almost circular shell of the HII region N49 \citep{WN49}}
\label{putaringonit}
\end{figure}

Beyond the uncertainties of the methods used, some of the underlying assumptions could also be responsible for this behaviour. First, the derivation of $K_\mathrm{e}$ for Fig. \ref{vlsrvske} and of the distance for the red clump method requires the knowledge of $A_\mathrm{J}/A_\mathrm{K}$. This coefficient depends mostly on the extinction law assumed. It is known that in the inner Milky Way this law seems to deviate from the standard values assumed by \citet{RL85} \citep[see for example][]{Ni06, Ni09}. All these works point towards a flattening of the extinction law, leading to more transparency in the $K_\mathrm{S}$ band, changing $A_J/A_K$ from the traditional $2.6$ to values near $6.0$ \citep{Ni06}. RSGC3 seems to have blown most of the molecular gas in its surroundings \citep{A09}, while Alicante 10 lies beyond some of this material and close to an active formation region (Fig. \ref{putaringonit}). This difference in environment could result in different values for $A_J/A_K$. The presence of this formation region and the gas surrounding it could have another effect. Because the evolution of $E(J-K_\mathrm{S})$ with $d$ from Fig. \ref{dvsejk} is measured over a region with $r=10'$ around Alicante 10, and the cluster itself occupies only $10\%$ of this area (roughly $6'\times6'$, see Fig. \ref{rgb}), the effect of this high-extinction region could be smeared out, tipping the $E(J-K_\mathrm{S})~\mathrm{vs}~d$ curve towards higher colour excesses.

Also, the long Galactic bar \citep{CCGF08} touches the Galactic arms/molecular ring in the vicinity of RSGC3. The potential and the orbits of a bar like this one are very complex, particularly so near its ends \citep[][e.g.]{At09}. One of the obvious effects would be a break in the circularity of the orbits assumed in Fig. \ref{vlsrvske}, and a difference between photometrically and dynamically derived distances should appear naturally. Sadly, there are no studies or estimations of these effects over other populations to compare our results, and so attributing the results of Fig. \ref{vlsrvske} and the discrepancy in the estimations of the distance of the effect of the bar only remains a possibility.

As can be seen in Fig. \ref{dvsejk} there is a rather steep increase in the extinction beyond $6~\mathrm{kpc}$. This behaviour it is also visible for RSGC3 (N11), and puts a strict upper limit on the distance to the sources. Since it is a more coherent estimate and requires less assumptions than the one derived from the values of $v_\mathrm{LSR}$, we opt to retain this estimate, concluding that both Alicante 10 and RSGC3 lie roughly at the same distance, $6~\mathrm{kpc}$.

Beyond its effect on the distance determination, this velocity difference between the two clusters is compatible with their being part of the same association. Similar dispersions have been observed in the molecular complex W43. This molecular cloud, already forming stars, sits just $\sim1^\circ$ away from RSGC3. The observed velocity range of its gas is $80-110~\mathrm{km\;s^{-1}}$, and while physically connected by material, W43-Main and W43-South, the two main clumps, peak at $\sim90~\mathrm{km\;s^{-1}}$ and $\sim100~\mathrm{km\;s^{-1}}$ respectively \citep{NG11}.

Based on this, we conclude that the populations of RSGC3 and Alicante 10 are similar, although the population from the latter seems to be slightly fainter,  hence might be older. It should be noted that this cluster lies near an active HII region, N49 \citep{WN49}. In fact, the RSG Near07 is very close to this region and has a velocity of $104~\mathrm{km\;s^{-1}}$, compatible with that of N49. If this turns out to be a real association would point to an age spread within the RSGC3 complex, although towards younger populations.

\subsection{Velocity outliers}

Even taking the natural dispersion in $v_{\mathrm{LSR}}$ and $K_\mathrm{e}$ into account, there are stars in Fig. \ref{vlsrvske} that deviate a lot from the bulk of the RSGC3 and Alicante 10 members. While some of them, like Halo01 and F1S04, are compatible in $K_{\mathrm{e}}$ and  $v_{\mathrm{LSR}}$ with the rotation curve (these two sources appear to be near the tangent point), other stars seem to deviate a lot from what this curve predicts. As has been stated previously, the interstellar extinction increases dramatically beyond $6~\mathrm{kpc}$, so it is very unlikely that these sources are at the far distance predicted by the rotation curve. Also, to be at a near distance, they should have $M_\mathrm{K}>-8$, far below any observed RSG in the Magellanic Clouds \citep{W83}. These stars are listed in Table \ref{tab3}, and we discuss them here in detail.

\begin{itemize}
\item[-] F1S01 and Halo02: The former has luminosity class Ib/II, while the latter shows $EW_{\mathrm{CO}}=-24.6~\AA{}$. This value falls in the lower range for RSGs, and might indicate a relatively high surface gravity, also implying classes Ib or II or a possible high-mass AGB star, which is a classification compatible with their later stellar type. This could point to a lower $M_\mathrm{K}$, and although both stars have very different $E(J-K_\mathrm{S})$, their velocities are compatible with being in the Galactic disk, and they could belong to the near Scutum-Crux arm.
\item[-] F1S10 and 2mass07: These stars have $v_{\mathrm{LSR}}$ and $K_\mathrm{e}$ compatible with being in the far side of the disk, yet this seems highly unlikely. F1S10 is a class Ib/II source, and it also has a very low colour excess, so it might be a foreground object, much like F1S01 or Halo02. The case of 2mass07 is a bit more complex. While it has $EW_{\mathrm{CO}}=-24.8~\AA{}$ and, as said before, might be a class Ib/II source, to be a foreground RSG it would need to have $M_\mathrm{K}\approx -6$, which is typical of a M3 giant, but these stars, as can be seen in Table \ref{tab1}, have lower $EW_{\mathrm{CO}}$. Although it could be a carbon-rich giant (with enhanced CO bands), its spectrum lacks all the features of those stars, such as deep CN absorptions. On the other hand, it has $E(J-K_\mathrm{S})$ too low to be on the far disk, and based on this indicator should be in the vicinity of Alicante 10. Its velocity is consistent with what is expected for the $3~\mathrm{kpc}$ arm around l=$29^\circ$, with the parametrization of \cite{Da08}. If this association should be true, it is very likely that this star is a RSG trapped in the potential of the bar, hence having non-circular motion incompatible with the rotation curves used in Fig. \ref{vlsrvske}.
\end{itemize}

\begin{table}
\caption{\label{tab3}Data of sources with peculiar dynamical behaviour.}
\centering
\begin{tabular}{ccccc} 
\hline\hline
ID&$K_\mathrm{S}$\tablefootmark{1} & $E(J-K_\mathrm{S})$ & $v_{\mathrm{LSR}}$  & Spectral\\
 & & & ($\mathrm{kms^{-1}}$) & type\\
\hline 
F1S01\tablefootmark{2}  &$5.71\pm0.02$&0.55&47&M7Ib/II\\   
F1S10\tablefootmark{2}  &$5.94\pm0.02$&0.95&70&M5Ib/II\\
Halo02 &$7.07\pm0.02$&2.66&42&M0I\\
2mass07&$8.51\pm0.02$&2.28&66&M0I\\
\hline
\end{tabular}
\tablefoot{
\tablefoottext{1}{$K_\mathrm{S}$ magnitudes from 2MASS.}
\tablefoottext{2}{Data taken from N11 with updated $v_{\mathrm{LSR}}$.}
}
\end{table}

\section{Conclusions}
We have conducted a study of a prospective cluster in the neighbourhood of RSGC3. For a number of photometrically selected candidates, spectroscopically based spectral types were obtained, finding that the vast majority of them are red supergiants. Our main results are as follows:
\begin{itemize}
\item[-] We confirm the existence of a new RSG cluster in the vicinity of RSGC3, named Alicante 10.
\item[-] This cluster has eight confirmed RSGs, implying a tentative initial mass well in excess of $10~000~M_{\odot}$.
\item[-] Although this association needs further study, based on its velocity and distance, Alicante 10 seems to be part of the RSGC3 complex. With the new confirmed RSGs, the census of these stars in the area grows to 35, confirming RSGC3 as one of the most massive events of stellar formation in the Galaxy.
\end{itemize}

Further spectroscopic and photometric studies of the area will help us understand the nature of this association, particularly its age, distance, and stellar content. Another issue that needs to be tackled is the dynamical complexity of the area, as basic parameters of the long bar, such as its pattern speed or the longitude of its axis, have a strong effect on it but are still poorly constrained.

\begin{acknowledgements}
      This research is partially supported by the Spanish Ministerio de Ciencia e Innovaci\'on (MCINN) under  AYA2010-21697-C05-5 and the Consolider-Ingenio 2010 programme grant CSD2006-00070: First Science with the GTC (http://www.iac.es/consolider-ingenio-gtc).
      This publication made use of data products from the Two Micron All Sky Survey, which is a joint project of the University of Massachusetts and the Infrared Processing and Analysis centre/California Institute of Technology, funded by the National Aeronautics and Space Administration and the National Science Foundation. This research made use of Aladin.
      
      The DENIS project was partly funded by the SCIENCE and the HCM plans of the European Commission under grants CT920791 and CT940627. It is supported by INSU, MEN and CNRS in France, by the State of Baden-W\"urttemberg in Germany, by DGICYT in Spain, by CNR in Italy, by FFwFBWF in Austria, by FAPESP in Brazil, by OTKA grants F-4239 and F-013990 in Hungary, and by the ESO C\&EE grant A-04-046.

Jean Claude Renault from IAP was the Project Manager.  Observations were carried out thanks to the contribution of numerous students and young 
scientists from all the involved institutes, under the supervision of  P. Fouqu\'e, survey astronomer resident in Chile. 

The UKIDSS project is defined in \citet{UK1}. UKIDSS uses the UKIRT Wide Field Camera \citep{UK2}. The photometric system is described in \citet{UK3}, and the calibration is described in \citet{UK4}. The pipeline processing and science archive are described in Irwin et al (2009, in prep) and \cite{UK5}.

IRAF is distributed by the National Optical Astronomy Observatories, which are operated by the Association of Universities for Research in Astronomy, Inc., under cooperative agreement with the National Science Foundation.
\end{acknowledgements}

\bibliographystyle{aa} % style aa.bst
\bibliography{al10_final} % your references Yourfile.bib

\begin{appendix} 
\section{Intrinsic infrared colour calibration for the red supergiants}
\label{JK0}
To derive the colour excesses for RSGs, an estimation of $E(J-K_\mathrm{S})$ is necessary. For these late type stars, the main reference sources are four: \citet{Gr04}, \cite{Ku05}, \citet{L05}, and \citet{Stra09}. While in the first work the authors centre their study over the supergiant population of the Magellanic clouds, the other three deal with stars located in the disk of the Milky Way. \citet{Stra09} and \citet{Ku05} provide intrinsic colours derived strictly from observations (once extinction is removed, which in itself needs the assumption of a reddening law).

Of these two, \citet{Stra09} compile colours for several giants of known spectral type, while in \cite{Ku05} there is no explicit relation between spectral type and $(J-K)_0$, as the authors classify their stars on a temperature scale. We transform these into spectral types using the derived temperatures \citet{Gr04} and \citet{L05} where available, and using the data from \cite{Ce01} everywhere else (marked with an asterisk in Table \ref{jktab}).

The other two works rely on synthetic colours calculated with the best-fitting models to a series of spectra. \citet{Gr04} derives  $(J-K)_0$ directly from these fits while \citet{L05} establishes a generic relation between temperature and intrinsic colour for the whole synthetic spectra database that can be combined with their own temperature determination to produce the desired $(J-K)_0$.

All the values in Table \ref{jktab} are transformed to the 2MASS photometric system using the updated relations from \citet{Ca01}\footnote{Available at http://www.astro.caltech.edu/~jmc/2mass/v3/transformations/}.

Even with these differences in methodology, it can be seen in Table \ref{jktab} that the values for  $(J-K_\mathrm {S})_0$ come in good agreement one with another, and yet some differences are obvious. For the earlier types there are are significant deviations beyond the typical error in the colour \citep[$\pm0.05~\mathrm{mag}$, ][]{Stra09}, which disappear at K5. Although we opted to combine the colours using a simple mean, it should be noted that other works point towards lower values of $(J-K)_0$ around the red clump \cite[typically spectral types K0 to K2, see][and references therein]{CCGF08}. Also, the photometry from \cite{Gr04} seems to be offset from the other two works by $0.04~\mathrm{mag}$. This difference can have two main sources: the authors combine 2MASS and DENIS data, the cross-calibration has a typical error of $0.05~\mathrm{mag}$, and they choose the reddening law of \citet{Dr03}, which produces a slightly different ratio of $E(J-K_\mathrm{S})$ to $A_\mathrm{K}$. When this work was the only source for $(J-K_\mathrm {S})_0$, this offset was subtracted. 
\begin{table*}
\caption{\label{jktab}Temperatures and intrinsic colour $(J-K_\mathrm {S})_0$ for the red supergiants}
\centering
\begin{tabular}{ccccccc|c} 
\hline\hline
&\multicolumn{2}{c}{Gr04\tablefootmark{1}}&\multicolumn{2}{c}{Le05\tablefootmark{3}}&St09\tablefootmark{4} & K05\tablefootmark{2} & Assumed\\
Type&$T_{\mathrm{eff}}$ & $(J-K_\mathrm {S})_0$&$T_{\mathrm{eff}}$ & $(J-K_\mathrm {S})_0$ & $(J-K_\mathrm {S})_0$ & $(J-K_\mathrm {S})_0$ & $(J-K_\mathrm {S})_0$\\
 &(K)&(mag)&(K)&(mag)&(mag)&(mag)&(mag) \\
\hline 
G5&---&---&---&---& 0.49& 0.47\tablefootmark{*}&0.48\\
G8&---&---&---&---& 0.56& 0.59\tablefootmark{*}&0.58\\
K0&---&---&---&---& 0.62& 0.67\tablefootmark{*}&0.65\\
K1&---&---& 4100& 0.86& 0.68& 0.88&0.87\\
K1.5&---&---& 4100& 0.86&---& 0.88&0.87\\
K2&---&---& 4015& 0.90& 0.75& 0.91&0.91\\
K3&---&---& 4015& 0.90& 0.82& 0.91&0.91\\
K4&---&---&---&---& 0.89&0.83\tablefootmark{*}&0.86\\
K5&---&---& 3840& 1.00& 0.96& 0.99&0.98\\
M0& 3850& 1.07& 3790& 1.03& 1.02&1.00&1.03\\
M1&---&---& 3745& 1.05& 1.06& 1.03&1.05\\
M1.5&---&---& 3710& 1.07&---& 1.05&1.06\\
M2&---&---& 3660& 1.10& 1.10& 1.07&1.09\\
M2.5&---&---& 3615& 1.12&---& 1.09&1.11\\
M3& 3550& 1.17& 3605& 1.13& 1.13& 1.13&1.13\\
M3.5&---&---& 3550& 1.16&---& 1.12&1.14\\
M4&---&---& 3535& 1.17& 1.17& 1.13&1.16\\
M4.5&---&---& 3535& 1.17&---& 1.13&1.15\\
M5& 3397& 1.26& 3450& 1.21& 1.21&---&1.21\\
M7& 3129& 1.36&---&---&---&---&1.32\\
M8& 2890& 1.36&---&---&---&---&1.32\\
M10& 2500& 1.34&---&---&---&---&1.30\\
\hline
\end{tabular}
\tablefoot{
\tablefoottext{1}{\citet{Gr04}}
\tablefoottext{2}{\citet{Ku05}}
\tablefoottext{3}{\citet{L05}}
\tablefoottext{4}{\citet{Stra09}}\\
\tablefoottext{*}{Derived using the temperature scale from \citet{Ce01}}
}
\end{table*}
\end{appendix}
\end{document}